\documentclass[conference]{IEEEtran}
\usepackage{epsfig}
\usepackage{graphics}
\usepackage{graphicx}
\usepackage{amsmath}
\usepackage{amssymb}
\usepackage{cite}
\usepackage{cases}
\usepackage{balance}
\usepackage{acronym}
\usepackage{algorithmic}
\usepackage[ruled,vlined]{algorithm2e}
\usepackage{tikz}
\usepackage{courier}

\usepackage{siunitx}
\DeclareSIUnit{\atm}{atm}
\DeclareSIUnit{\kWh}{kWh}
\DeclareSIUnit{\Ah}{Ah}
\DeclareSIUnit \voltampere { VA }
\DeclareSIUnit \wh { Wh }
\DeclareSIUnit \rad { rad }

\def \acc {\`}

\newtheorem{lemma}{Lemma}
\newtheorem{theorem}{Theorem}

\newtheorem{definition}{Definition}

\acrodef{pmu}[PMU]{Phasor Measurement Unit}
\acrodef{wls}[WLS]{Weighted Least Squares}
\acrodef{wmr}[WMR]{Weighted Measurement Residual}
\acrodef{fdla}[FDLA]{Fault Detection and Localization Algorithm}
\acrodef{fdl}[FDL]{fault detection and localization}
\acrodef{ufc}[UFC]{Unlocalizable Fault Cluster}
\acrodef{der}[DER]{Distributed Energy Resources}
\acrodef{adn}[ADN]{Active Distribution Network}
\acrodef{dn}[DN]{Distribution Network}
\acrodef{mv}[MV]{Medium Voltage}
\acrodef{se}[SE]{State Estimator}
\acrodef{ba}[BA]{Basic Algorithm}
\acrodef{zsic}[ZSIC]{zero-sequence injected current}
\acrodef{feg}[fEG]{fake extended grid}
\acrodef{ufc2}[UFC*]{re-defined Unlocalizable Fault Cluster}
\acrodef{opa}[OPA]{Optimal Positioning Algorithm}

\IEEEoverridecommandlockouts

\newcommand\copyrighttext{%
  \footnotesize
  \centering\copyright~2021 IEEE. Personal use of this material is permitted. Permission from IEEE must be obtained for all other uses, in any current or future media, including reprinting/republishing this material for advertising or promotional purposes, creating new collective works, for resale or redistribution to servers or lists, or reuse of any copyrighted component of this work in other works.\\
  DOI: 10.1109/PowerTech46648.2021.9494871.}
\newcommand\copyrightnotice{%
\begin{tikzpicture}[remember picture,overlay]
\node[anchor=south,yshift=0pt] at (current page.south) {\setlength{\fboxrule}{0pt}\fbox{\parbox{\dimexpr\textwidth-\fboxsep-\fboxrule\relax}{\copyrighttext}}};
\end{tikzpicture}%
}

\begin{document}

%Title of the paper
\title{Optimal Positioning of PMUs for Fault Detection and Localization in Active Distribution Networks\thanks{This work has been financed by the Research Fund for the Italian Electrical System in compliance with the Decree of Minister of Economic Development April 16, 2018.}}
%Authors

\author{\IEEEauthorblockN{F. Conte, B. Gabriele, G.-P. Schiapparelli, F. Silvestro}
\IEEEauthorblockA{\textit{DITEN - Universit\acc{a} degli Studi di Genova} \\
Genova, Italy \\
fr.conte@unige.it}
\and
\IEEEauthorblockN{C. Bossi, M. Cabiati}
\IEEEauthorblockA{\textit{Ricerca sul Sistema Energetico - RSE S.p.A.} \\
Milano, Italy \\
claudio.bossi@rse-web.it}
}

\IEEEaftertitletext{\copyrightnotice\vspace{1.1\baselineskip}}
\maketitle

%Abstract
\IEEEpubidadjcol
\begin{abstract}
This paper considers the problem of fault detection and localization in active distribution networks using \acp{pmu}. The proposed algorithm consists in computing a set of weighted least squares state estimates whose results are used to detect, characterize and localize the occurrence of a fault. Moreover, a criteria to minimize the number of \acp{pmu} required to correctly perform the proposed algorithm is defined. Such a criteria, based on system observability conditions, allows the design of an optimization problem to set the positions of \acp{pmu} along the grid, in order to get the desired fault localization resolution. The performances of the strategy are tested via simulations on a benchmark distribution system.
\end{abstract}


%%%%%%%%%%%%%%%%%%%%%%%%%%%%%%%%%%%%%%%%%%%%%%%%%%%%%%%%%%%%%%%%%%%
\section{Introduction} \label{Introduction}
%%%%%%%%%%%%%%%%%%%%%%%%%%%%%%%%%%%%%%%%%%%%%%%%%%%%%%%%%%%%%%%%%%%
Nowadays, \acp{dn} are operated in a very different way with respect to their original design~\cite{bak2015protection,Conte:2019}. The massive use of \acp{der} causes non-unidirectional power flows with a dramatic impact on the protection scheme behaviour.
The main issues can be classified as \cite{telukunta2017protection}: blinding of protection, false tripping or sympathetic tripping, islanding problems, loss of coordination, auto-recloser problems.
Literature proposes several protection schemes to address these problems. They are generally based on: limitation of \ac{der} capacity \cite{chaitusaney2008prevention}, use of distance protections and reconfiguration strategies \cite{ sinclair2013distance}, in line fault current limiters \cite{yamaguchi2008current}, adaptive protection schemes. Specifically, the latter assumes that relays are able to communicate each others.

Recent studies investigate the possibility of merging protection devices, distributed measurements and communication systems, in order to asses in real-time the system state and take decisions based on a complete view of the grid operating conditions. Such an approach opens to a plethora of applications both in operation and planning \emph{e.g.} state estimation \cite{milano:2016, conte2020assessment,Adinolfi:2013}, system and load modelling, event detection and localization, optimal positioning of measurement devices and switches \cite{liu2012trade}.

The focus of this paper is on \ac{fdl} in \acp{adn} using the measurements provided by \acp{pmu}. Recent literature proposes several interesting solutions. 
In \cite{pignati2016fault}, \ac{fdl} is realized for \acp{adn} by placing \acp{pmu} at each node of grid, allowing the localization of the faulted line. In \cite{farajollahi2017location}, \ac{fdl} is realized for radial passive \acp{dn}, allowing the localization of a fault between a couple of \acp{pmu}. This method requires pseudo-measurements that introduce uncertainty in the provided results. The technique developed in \cite{ardakanian2017event} allows \ac{fdl} in radial passive \acp{dn}. In this case, \acp{pmu} are placed at each node of the grid. In \cite{majidi2017new}, \ac{fdl} is realized for \acp{adn} by placing \acp{pmu} according to a suitable positioning criteria. With this method, localization results are approximated. In \cite{gholami2019detecting} \ac{fdl} is carried out for \acp{adn} by placing \acp{pmu} only to nodes with \acp{der}. Here pseudo-measurements are used and localization regards the identification of the faulted portion of the grid. Reference \cite{Jamei:2019} proposes a method for \ac{fdl} for \acp{adn} by placing \acp{pmu} at the point of connection with the main grid, at all nodes with \acp{der}, and at a set of nodes defined by optimal positioning. In this case, localization results are approximated, but estimation errors are characterized by a probabilistic model. 

The present paper proposes a method of \ac{fdl} in radial \ac{adn}. The approach is similar to the one of \cite{pignati2016fault}, where a set of \ac{wls} \acp{se} are performed considering all possible network topologies for each fault location (each line). Differently from \cite{pignati2016fault}, the objective is to perform \ac{fdl} without placing \acp{pmu} at each node, but looking for the minimum necessary number of such devices. With this aim, a \acp{pmu} \ac{opa} is developed based on observability conditions, suitably defined according to the properties of the adopted estimation procedure. The result of the \ac{opa} is the partition of the grid into connected lines clusters, where the proposed \ac{fdla} is able to localize the fault. The advantage of the method is that the \ac{opa} can be customized to obtain a desired localization resolution, in terms of number and composition of the lines clusters.

The paper is organized as follows: Section~\ref{sec:ProblemFormulation} reports the general problem formulation; Section~\ref{sec:wls} introduces the basic state-estimation based method; Section~\ref{sec:observability} defines observability and fault localizability conditions; Section~\ref{sec:FaultDetectionLocalization} provides the \ac{fdla} and the \acp{pmu}-\ac{opa}; Section~\ref{sec:CaseStudy} describes the case study; Section~\ref{sec:Results} presents simulation results; Section~\ref{sec:Conclusions} reports the conclusions of the paper.

\vspace{10pt}
\textit{Notation.} $\dot{\mathbf{U}}^{abc}$ is a triplet vector $\begin{bmatrix} \dot{U}^{a} & \dot{U}^{b} & \dot{U}^{c} \end{bmatrix}$. Real and imaginary components of a generic phasorial vector $\dot{U}$ are indicated as $U_{re}$ and $U_{im}$. $b_i$ indicates the $i$-th node of the grid; $\rho(b_i)$ is the degree of $b_i$, defined as the number of nodes connected with $b_i$.

%%%%%%%%%%%%%%%%%%%%%%%%%%%%%%%%%%%%%%%%%%%%%%%%%%%%%%%%
\section{Problem Formulation} \label{sec:ProblemFormulation}
%%%%%%%%%%%%%%%%%%%%%%%%%%%%%%%%%%%%%%%%%%%%%%%%%%%%%%%%
The purpose of this work is to detect and locate a short-circuit event whenever it occurs in a \ac{mv} distribution grid. The following assumption are made: \emph{1)} distribution network is radial and composed by $n$ nodes and $m$ lines; \emph{2)} the network admittance matrix $Y$ is known; \emph{3)} $d<n$ nodes are monitored via \acp{pmu}, and $n-d$ nodes are not; $\mathcal{I}_m^d$ is the set of the indices of the $d$ monitored nodes;  \emph{4)} at each monitored node $b_i$ ($i\in\mathcal{I}_m^d$), the measurement system acquires the measures of three phase-to-ground voltages $\dot{\mathbf{V}}_i^{abc}$ and three phase node injected currents $\dot{\mathbf{I}}_i^{abc}$.

Following these assumptions, state vector $x$ is defined as
\begin{align}
& x=\begin{bmatrix}   \mathbf{V}^{abc}_{1,re} & \cdots & \mathbf{V}^{abc}_{n,re} &  \mathbf{V}^{abc}_{1,im}& \cdots & \mathbf{V}^{abc}_{n,im} \end{bmatrix}^\top \in \mathbb{R}^N, 
\end{align}
with $N = 6n$, and measurements vector $z$ is defined as
\begin{equation}\label{eq:measurementVector}
z=[z_V,z_I]^\top \in \mathbb{R}^{D}, \quad D=12d,
\end{equation}
%with:
\begin{align}
	z_V & = 
	\begin{bmatrix} \cdots & \mathbf{V}^{abc}_{re,i} & \cdots &\mathbf{V}^{abc}_{im,i} & \cdots 
    \end{bmatrix},  \quad i \in \mathcal{I}_m^d, \label{eq:zV} \\
	z_I&=
	\begin{bmatrix}
	\cdots & 
    \mathbf{I}^{abc}_{re,i} & \cdots  &\mathbf{I}^{abc}_{im,i} & \cdots 
    \end{bmatrix}, \quad i \in \mathcal{I}_m^d. \label{eq:zI}
\end{align}

\acp{pmu} introduce noise in the measurements, supposed to be additive, zero-mean and with a known covariance matrix $R$. Refer to \cite{milano:2016} to analyze the validity of assuming that \acp{pmu} provide voltage and current in Cartesian coordinates.

The relation among $x$ and $z$ has the form:
\begin{equation}\label{eq:output equation}
    z = Hx + v
\end{equation}
where $H$ is a $D\times N$ matrix and $v$ is the measurement noise. 

Matrix $H$ has the form:
\begin{equation}
H = \begin{bmatrix}
    H_V \\
    H_I
\end{bmatrix}
\end{equation}
where $H_V\in\mathbb{R}^{6d \times N}$ relates the state vector $x$ with the component $z_V$ of $z$, and $H_I\in\mathbb{R}^{6d \times N}$ relates $x$ with the component $z_I$ of $z$. Therefore, $H_V$ is composed by zeros and ones to select the voltages directly measured by \acp{pmu}, $H_I$ is computed using the network admittance matrix $Y$, by removing from the following matrix
\begin{equation}
\mathcal{H}_I = \begin{bmatrix}
    Re\{Y\} & -Im\{Y\} \\
    Im\{Y\} & Re\{Y\} 
\end{bmatrix},
\end{equation}
the rows corresponding to non-monitored nodes.

Under the mentioned assumptions, the objectives of the paper are: \emph{1)} to develop a \acf{fdla} that, using the available \acp{pmu} measurements, is able to: \emph{1.a)} \textit{detect} the occurrence of a fault; \emph{1.b)} \textit{characterize} the fault, \textit{i.e.} state in which phases it has occurred; \emph{1.c)} \textit{localize} the fault; \emph{2)} to define the minimum number $d$ and the optimal positions of \acp{pmu} along the grid, that allow the \ac{fdla} to work correctly and obtain a desired fault localization resolution.

%%%%%%%%%%%%%%%%%%%%%%%%%%%%%%%%%%%%%%%%%%%%%%%%%%%%%%%%
\section{Basic Algorithm}\label{sec:wls}
%%%%%%%%%%%%%%%%%%%%%%%%%%%%%%%%%%%%%%%%%%%%%%%%%%%%%%%%
In this section, we provide the core algorithm of the general \ac{fdla}, which will be illustrated  in Section~\ref{sec:FaultDetectionLocalization}. The adopted approach is similar to the one proposed in \cite{pignati2016fault}. The idea is that a fault on a line results in a sudden addition of one \textit{virtual} node in the network, that is between two real nodes and absorbs the fault current. Therefore, $m+1$ parallel \ac{wls} \acp{se}  \cite{milano:2016} are realized, each returning, at any measurement time step, the estimate $\hat{x}^k$, $k=0,1,\ldots,m$. Estimate $\hat{x}^{0}$ is obtained by applying \ac{wls} without adding virtual nodes, as follows: 
\begin{equation}\label{eq:wls0}
\hat{x}^{0}=\left({H}^\top R^{-1} H\right)^{-1} {H}^\top R^{-1} z.
\end{equation}

Estimates $\hat{x}^{k}$, with $k>0$, are computed by applying the \ac{wls} equation to the network extended with a virtual node placed in the middle of line $k$. Specifically, for each $k=1,2,\ldots,m$: \emph{a)} the state vector $x$ is extended with the voltage of the virtual node $b_{n+1}$:
\begin{equation}
x^k=\begin{bmatrix}   \mathbf{V}^{abc}_{1,re} & \cdots & \mathbf{V}^{abc}_{n+1,re} & \mathbf{V}^{abc}_{1,im}& \cdots &  \mathbf{V}^{abc}_{n+1,im}\end{bmatrix}^\top; 
\end{equation}
\emph{b)} the admittance matrix of the extended grid $Y^k$ is computed and used to obtain the corresponding measurement matrix $H^k$; \emph{c)} estimate $\hat{x}^k$ is computed as:
\begin{equation}\label{eq:wlsk}
\hat{x}^{k}=\left({H^k}^\top R^{-1} H^k\right)^{-1} {H^k}^\top R^{-1} z.
\end{equation}

Each \ac{se} is associated to the \ac{wmr},
\begin{equation}\label{eq:wmr}
w^k=(z-H^k\hat{x}^k)^\top R^{-1}(z-H^k\hat{x}^k),
\end{equation}
to evaluate the estimate accuracy.

As in \cite{pignati2016fault} the idea is that, in a no-fault scenario, this process will return $m+1$ estimates with comparable \acp{wmr}, since all \acp{se} use a correct model of the grid topology, in which the virtual node will be suitably estimated to do not absorb any current. Differently, if a fault occurs for example on line $\alpha$, only the $\alpha$-th \acp{se} will be based on a topology close to real one. This implies that all \acp{wmr}, excepting for the $\alpha$-th, will sudden increase. In particular, the $0$-th \ac{wmr} will always increase after a fault, since the grid topology model used by the corresponding \ac{se} does not include any virtual node.       

Therefore, the \ac{ba} consists in the following steps: \emph{1)} compute the $m+1$ \ac{wls} estimates; \emph{2)} detect the fault by registering an anomalous variation of $w^0$; \emph{3)} localize the fault by selecting the minimum \ac{wmr} \emph{4)} characterize the fault by computing the estimated currents injected into the virtual node from the selected best estimate $\hat{x}^{\alpha}$ (using the corresponding admittance matrix: $Y^{\alpha}\hat{x}^{\alpha}$) and registering which of the three phase currents are different from zero.

Unfortunately, the \ac{ba} does not generally works in the hypothesis that not all nodes are monitored by \acp{pmu} (\textit{i.e.} $d<n$). First, \ac{wls} estimates \eqref{eq:wls0} and \eqref{eq:wlsk} cannot be implemented independently of the number and the position of \acp{pmu}, since this is possible only if the network \textit{observabilty} is kept \cite{Abur:2004}. Second,
even if estimates \eqref{eq:wls0}--\eqref{eq:wlsk} are computable, localization cannot correctly carried out just selecting the minimum \ac{wmr}. 

This fact is clarified in the following section, where, moreover, the conditions to keep the network observability with a reduced number of \acp{pmu} are provided.

%%%%%%%%%%%%%%%%%%%%%%%%%%%%%%%%%%%%%%%%%%%%%%%%%%%%%%%%%%%%%%%%%
\section{Grid observability and fault localizability}\label{sec:observability}
%%%%%%%%%%%%%%%%%%%%%%%%%%%%%%%%%%%%%%%%%%%%%%%%%%%%%%%%%%%%%%%% 
Following the definitions in \cite{Abur:2004}, an electrical network is observable if, in the no noise ideal case, it is possible to calculate the voltage at all nodes starting from the available measurements. With the formulation used in this paper, we have observability if and only if matrix $H$ is full row-rank. It is well known that, if this does not hold true, matrix $H^\top R^{-1}H$ in \eqref{eq:wls0} is not invertible and, thus, \ac{wls} cannot be used.      

Obviously, if all nodes are equipped with a \ac{pmu}, all nodal voltages are directly known ($H_V$ is the identity matrix in $\mathbb{R}^{6n}$), and the network is observable. However, our objective is to reduce as match as possible the number of \acp{pmu}. 

By applying simple electrotechnical computations it is easy to prove the following lemma.
\begin{lemma}\label{lem:1}
A radial grid with $n$ nodes and $m$ lines is observable if, for all $i=1,2,\ldots,n$:
\begin{itemize}
\item[a.] if $\rho(b_i)>1$, $b_i$ has at most one adjacent non-monitored node;
\item[b.] if $\rho(b_i)=1$ and $b_i$ is non-monitored, then $b_i$ is adjacent to a monitored node.
\end{itemize}
\end{lemma}

Lemma \ref{lem:1} gives us the conditions to allow \eqref{eq:wls0} to be implemented. However, we also need to implement estimates \eqref{eq:wlsk}. Therefore, the observability of the $m$ networks extended with a virtual node placed in the middle of each line is required. Indeed, this will imply that matrices $H^k$ are full row-rank and, thus, ${H^k}^\top R^{-1} {H^k}$ are invertible. The following theorem provides the conditions to satisfy such a requirement. 

\begin{theorem}\label{th:1}
Given a radial grid with $n$ nodes and $m$ lines, the extended grid with $n+1$ nodes and $m+1$ lines, obtained by adding the virtual node $b_{n+1}$ in the middle of one of the $m$ lines is observable if and only if, for all $i=1,2,\ldots,n$:
\begin{itemize}
\item[a.] whatever given a couple of adjacent nodes, at least one of the two is monitored;
\item[b.] if $\rho(b_i)=1$, $b_i$ is monitored.
\end{itemize}
\end{theorem}
\textit{Proof.}
By definition, $\rho(b_{n+1})=2$ and it is non-monitored. Thus, condition a. of Lemma~\ref{lem:1} implies that, by positioning the virtual node in the middle of any line of the grid, it has to be adjacent to a monitored node. This is verified if and only if condition a. of Theorem~\ref{th:1} holds true. Condition b. of Lemma~\ref{lem:1} is a necessary and sufficient for condition b. of Theorem~\ref{th:1}. Indeed, if the virtual node is located at a line incident to a node $b_i$ with degree $\rho(b_i)=1$, we have a non-monitored node (the virtual one) with a node with degree = 1. Thus, condition b. of Theorem~\ref{th:1} is verified only if condition b. of Lemma~\ref{lem:1} holds true, and \textit{viceversa}. The sufficient condition of Theorem~\ref{th:1} is, therefore, proved. The necessary condition can be proved by contradiction showing that there exist counterexamples when one of the two conditions are not satisfied. 

Roughly speaking, Theorem~\ref{th:1} states that the terminal nodes of each feeder must be equipped with a \ac{pmu} and there cannot be two adjacent nodes without \ac{pmu}.

With Theorem~\ref{th:1}, we have the conditions to make estimates \eqref{eq:wls0}--\eqref{eq:wlsk} implementable and correctly working.
Next Theorem~\ref{th:2} provides a key property of such estimates that makes a fault unlocalizable within specific portions of the grid by using the \ac{wls} estimates. Such grid portions are defined as follows.

\begin{definition}\label{def:1}
An \ac{ufc} is a connected portion of the grid, defined as a subset of lines $\mathcal{C}\in\{1,2,\ldots,m\}$, such that:
\begin{itemize}
    \item if $\mathcal{C}$ is composed by one single line, then this line connects two monitored nodes (single-line \ac{ufc});
    \item if $\mathcal{C}$ is composed by more than one line, then: \textit{a)} it is connected to the rest of the grid through a non-monitored node; and \textit{b)} whatever given one line in $\mathcal{C}$, it connects a monitored node to a non-monitored one.
\end{itemize}
\end{definition}

\begin{theorem}\label{th:2}
Consider a radial grid with $n$ nodes and $m$ lines and suppose that conditions a. and b. of Theorem~\ref{th:1} are satisfied. Whatever given an \ac{ufc} $\mathcal{C}$ composed by more than one line, and two estimates $\hat{x}^{i}$ and $\hat{x}^{\ell}$, computed by \eqref{eq:wlsk}, associated to virtual nodes placed on two lines belonging to $\mathcal{C}$, then, for all $z\in\mathbb{R}^{D}$, $w^{i}=w^{\ell}$.
\end{theorem}

We cannot provide a formal proof of Theorem~\ref{th:2} in this paper for space lacking. %However, from \eqref{eq:wlsk} and \eqref{eq:wmr}, it follows that, for all $k$, 
%\begin{equation}\label{eq:proofth2}
%w^{k} = z^\top\left(I-G^{k}\right)^{\top}  R^{-1} \left(I-G^{k}\right) z
%\end{equation}
%where  $G^{k}={H^k}({H^k}^\top R^{-1} H^k)^{-1} {H^k}^\top R^{-1}$. It can be shown that, whatever given an \ac{ufc} $\mathcal{C}$, matrices $G^{i}$ and $G^{\ell}$ related to virtual nodes placed on two lines belonging to $\mathcal{C}$, are identical. This proves Theorem~\ref{th:2} since from \eqref{eq:proofth2} it follows that for all $z\in\mathbb{R}^D$, $G^{i}=G^{\ell}\Rightarrow w^{i}=w^{\ell}$.    

Theorem~\ref{th:2} implies that, using the \ac{ba} introduced in Section~\ref{sec:wls}, we will obtain a set of grid clusters associated to \ac{wls} estimates with identical \acp{wmr}. Therefore, an idea could be to select the minimum \ac{wmr} to discover in which \ac{ufc} the fault has occurred. Unfortunately, differently on what happens when all nodes are monitored \cite{pignati2016fault}, in the present case, putting a virtual node in the middle of one line within the correct \ac{ufc} does not necessarily result in the best representation of the faulted grid, excepting for the unlikely case of occurrence of the fault at the exact half of the line. 

In particular, this happens when the fault occurs within a single-line \ac{ufc}. We verified that, in these cases, if the fault occurs sufficiently far from the middle of the line, the \ac{wmr} of any adjacent \ac{ufc} could result to be smaller than the one associated to the right one. This can be explained by the fact that, in this scenario, both the \ac{wls} \acp{se} associated to the right and the wrong adjacent \acp{ufc} use an approximated model of the grid topology. However, differently from the right single-line \ac{ufc}, the wrong adjacent \acp{ufc} could include non-monitored nodes close to the real fault location, which gives to the estimation process a degree of freedom that allows to compensate the modelling error. 

To fix this problem, the \acp{se} associated to single-line \acp{ufc} should be put in condition to work with the same degrees of freedom of those associated to the adjacent \acp{ufc}. The solution is to add \textit{fake nodes} in the middle of all single-line \acp{ufc}, which, by Definition~\ref{def:1}, are all lines between two monitored nodes, and apply the approach to the so obtained extended grid. More formally, we first define this last as \ac{feg}. Then, we can apply Definition~\ref{def:1} to the \ac{feg} to obtain the relevant \acp{ufc}, which, in this case, cannot be single-line. By Theorem~\ref{th:2} there will be a unique \ac{wmr} associated to each of these \acp{ufc}, but now the one corresponding to the right one will return the minimum \ac{wmr}.

Since we are not interested on the definition of line clusters that include lines incident to fake nodes, we define as follows, the \ac{ufc2}, which coincide with the \acp{ufc} of the \ac{feg} without fake nodes.     

\begin{definition}\label{def:2}
An \ac{ufc2} is a connected portion of the grid, defined as a subset of lines $\mathcal{C}\in\{1,2,\ldots,m\}$ that is connected to the rest of the grid through a non-monitored node with degree $>2$.
\end{definition}

%%%%%%%%%%%%%%%%%%%%%%%%%%%%%%%%%%%%%%%%%%%%%%%%%%%%
\section{Fault Detection and Localization Algorithm and Optimal Positioning of PMUs} \label{sec:FaultDetectionLocalization}
%%%%%%%%%%%%%%%%%%%%%%%%%%%%%%%%%%%%%%%%%%%%%%%%%%%%
In this section, we finally provide: the \ac{fdla}, which is able to detect, characterize and locate the occurrence of a fault within one of the grid \acp{ufc2}, and the \acp{pmu} \acf{opa} by which it is possible to choice the number and the positions of \acp{pmu} along the grid nodes, in order to allow the \ac{fdla} to work correctly and get a desired fault localization resolution.

\subsection{FDLA}
The general pseudo-code of the \ac{fdla} is reported in Algorithm~\ref{alg:1}. 
First of all, the number and the composition of the grid \acp{ufc2} is determined by function \texttt{\small getUFC}. This function is designed according to Definition~\ref{def:2}, which states that \acp{ufc2} are separated by non-monitored nodes with degree $>2$. Therefore, using standard graph analysis algorithms, and given the grid topology and the set of monitored nodes $\mathcal{I}^d_m$, \texttt{\small getUFC} returns the number $r<m$ and the composition of the \acp{ufc2}, $\{\mathcal{C}^l\}$, $l=1,2,\ldots,r$, where $\mathcal{C}^l\in\{1,2,\ldots,m\}$. 

Then, at each measurement time step $t$, one state estimate associated to each \acp{ufc2} $x^{l}(t)$ and one associated to the grid without virtual nodes, the $0$-th $\hat{x}^{0}$, and the relevant \acp{wmr} are computed. Then, the variation of $w^0$ is checked: if it crosses the threshold $th_w$, a fault is detected. 
Then, the faulted \ac{ufc2} is determined by selecting the \ac{wmr} $w^{\alpha}$ that realizes the minimum variation with respect the average value $\mu_w^l$ computed over the preceding samples.
Function {\texttt{\small getCluster}}$\left(\alpha\right)$ returns the \ac{ufc2} $\mathcal{C}^*$ of belonging of line $\alpha$. Finally, function \texttt{\small characterizeFault}($\hat{x}^{\alpha}$) returns the \textit{Fault Type}, \textit{i.e.} the faulted phases, using step \emph{4)} of the \ac{ba}.

%%%%%%%%%%%%%%%%%%%%%%%%%%%%%%%%%%%%%%%%%%%%%%%%%%%%%
\subsection{PMUs-OPA}\label{sec:OptimalPositioning}
%%%%%%%%%%%%%%%%%%%%%%%%%%%%%%%%%%%%%%%%%%%%%%%%%%%%%
The localization resolution of the \ac{fdla} depends on the number $r$ of \acp{ufc2}. According to Definition~\ref{def:2}, $r$ is augmented if fork nodes (nodes with degree $>2$) are forced to be non-monitored as much as possible. Therefore, using also Theorem~\ref{th:1} to assure grid observabilty, we can define the \acp{pmu}-\ac{opa}, which consists in the solution of the following mixed-integer optimization problem:   
\begin{align}
	& \min_{\gamma} c^\top \gamma \label{eq:positioningObjective} \\
	& s.t.
    \begin{cases}
            A\gamma\geq{f} \label{eq:positioningConstraints} \\ 
	        \gamma_i=1 &\mbox{if } \rho(b_i)=1 \quad \forall i=1,2,\ldots,n
    \end{cases}
\end{align}
where: $\gamma=[\gamma_1 \ \gamma_2 \ \cdots \gamma_n]^\top$ is a vector of binary variables representing if the $i$-th node is monitored $(\gamma_i=1)$ or not $(\gamma_i=0)$;
matrix $A\in\mathbb{R}^{n \times n}$ describes the grid topology:
\begin{equation} \label{eq:incidenceMatrix}
    A_{i,k}= 
    \begin{cases}
        \rho(b_i) &\mbox{if } i=k \\
        1 &\mbox{if } (i\neq k) \mbox{ $\wedge$ ($b_i$ and $b_k$ connected)} \\
        0 & \mbox{otherwise}
    \end{cases}
\end{equation}
$f\in\mathbb{R}^n$ with $f_i=\rho(b_i)$; and $c\in\mathbb{R}^n$ collects cost function weights $c_i$.

Constraints in \eqref{eq:positioningConstraints} imposes conditions of Theorem~\ref{th:1}. Cost function weights $c_i$ can be all equal to one if the unique objective is to minimize the number $d$ of required \acp{pmu}, or defined as follows, if the objective is also to maximize the number $r$ of \acp{ufc2}:
\begin{equation} \label{eq:cost_function_option2}
    c_i =  \begin{cases}
        1 &\mbox{if } \rho(b_i)\leq 2 \\
        n\cdot \rho(b_i)  & \mbox{if } \rho(b_i)> 2 \\
    \end{cases}
\end{equation}
In this way, optimization will strongly penalize the positioning of \acp{pmu} on fork nodes.

Finally, if the user desires to distinguish two specific portions of the grid, for example to perform protections, we can state that they should be divided by a fork node $b_\kappa$. If this holds true, and $b_\kappa$ is not adjacent to another fork node, to apply \eqref{eq:cost_function_option2} is enough to obtain the desired grid portioning. Differently, if $b_\kappa$ is adjacent to other fork nodes, we need to modify \eqref{eq:cost_function_option2} by setting equal to one the weights $c_i$ corresponding to all the fork nodes adjacent to $b_\kappa$.

\begin{algorithm}[t] \label{alg:1}
 \caption{\ac{fdla}}
\SetAlgoLined
$[r,\{C^l\}]\leftarrow$  \texttt{\small getUFC}(\textit{Grid Topology}, $\mathcal{I}^d_m$)\;
\textbf{for all} measurement time step $t$\;
  collect \acp{pmu} measurements $z(t)$\;
  compute $\hat{x}^{l}(t)$ and $w^{l}(t)$, $0=1,2,\dots,r$\;
  \If{$|w^0(t)-w^0(t-1)|>th_w$}{
   $\rightarrow$ \textit{Fault Detected}\;
   $\alpha = \min_{l}\{w^{l}(t)-\mu_{w}^l\}$\;
   $\mathcal{C}^*$ = \texttt{\small getUFC}($\alpha$) $\rightarrow$ \textit{Faulted Cluster}\;
   \texttt{\small characterizeFault}($\hat{x}^{\alpha}$) $\rightarrow$ \textit{Fault Type}\;
   }
\end{algorithm}

%%%%%%%%%%%%%%%%%%%%%%%%%%%%%%%%%%%%%%%%%%%%%%%%%%%%%%%%
\section{Case Study} \label{sec:CaseStudy} 
\begin{figure}[t]
	\centering
	\includegraphics[width=0.8\columnwidth]{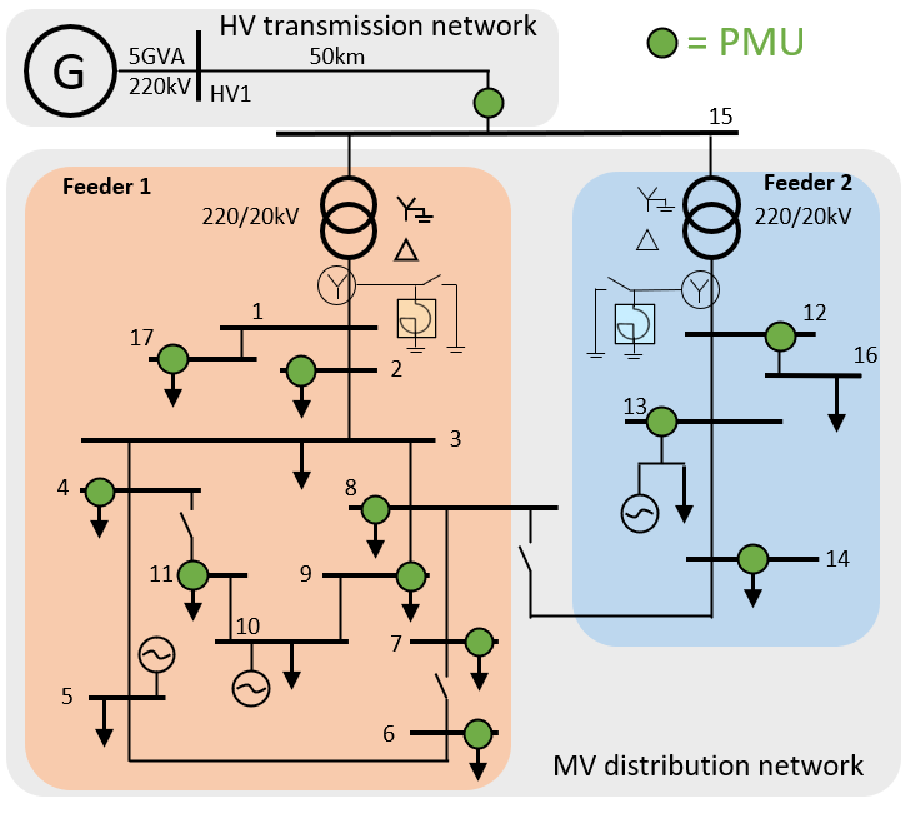}
    	\caption{\ac{adn} case study.}
	\label{fig:Grid}
	\vspace{-10pt}
\end{figure}

%%%%%%%%%%%%%%%%%%%%%%%%%%%%%%%%%%%%%%%%%%%%%%%%%%%%%%%%
The performances of the proposed solution are evaluated by simulations on a modified version of the Cigré benchmark \ac{mv} distribution network \cite{ramos2015benchmark}, shown in Fig.~\ref{fig:Grid}. There are $17$ nodes, divided into two feeders
starting from nodes 1 and 12, from which the neutral is earthed through grounding transformers. The grid has been implemented with the option of connecting the neutral directly to the ground or by Petersen coil. Loads are delta-connected and represent secondary substations. Finally, three \SI{400}{kVA} Diesel generators have been placed at nodes 5, 10 and 13.

\begin{figure}[t]
	\centering
	\includegraphics[width=0.85\columnwidth]{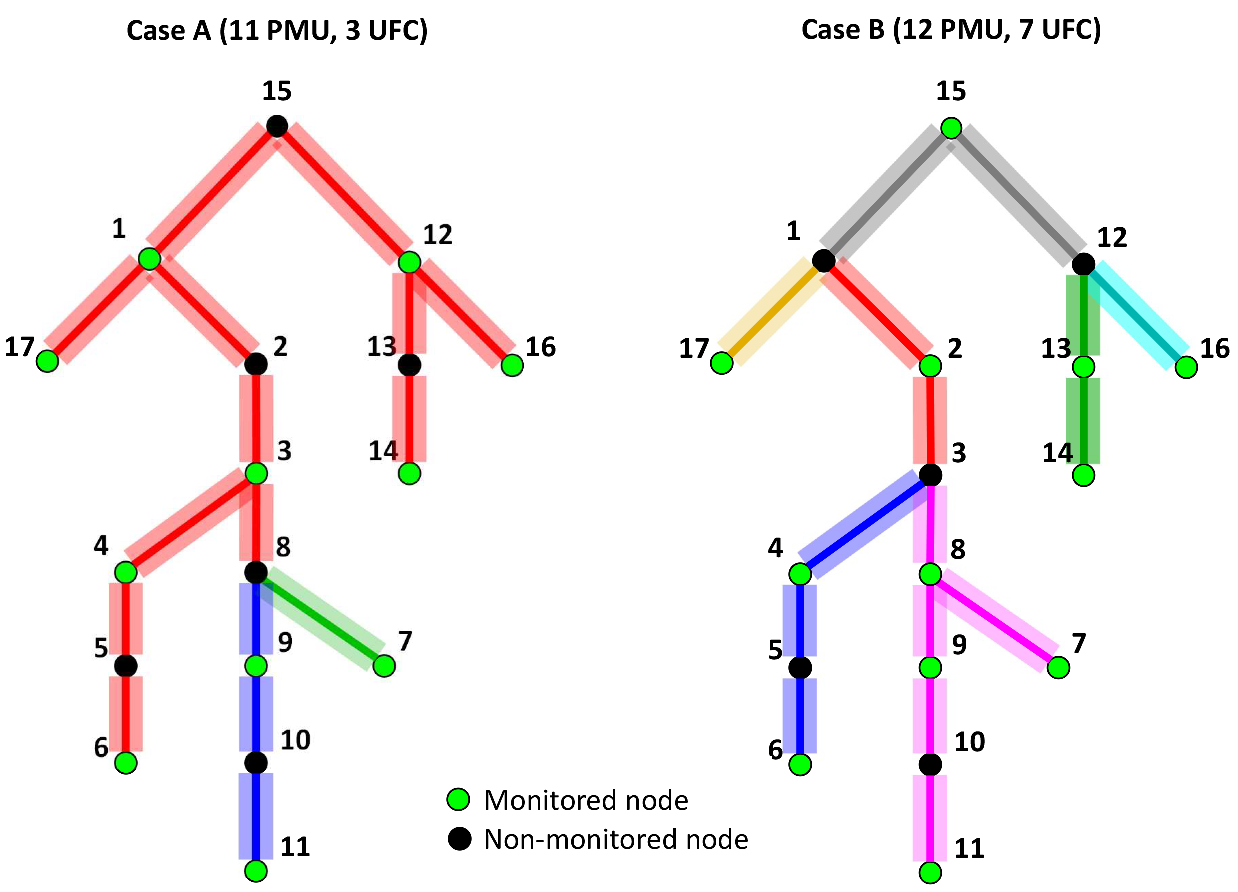}
    	\caption{\acp{pmu} optimal positioning. In Case~A (left) the number of $d$ \acp{pmu} is minimized. In Case~B (right) $d$ is minimized and the number $r$ of \acp{ufc2} is maximized. Lines are colored according to the composition of the \acp{ufc2}.}
	\label{fig:opa}
\end{figure}

Figure~\ref{fig:Grid} reports the positions of \acp{pmu}, defined using the \ac{opa}. Figure~\ref{fig:opa} compares the results of two versions of the optimization. In Case A, cost function weights are all equal to 1. Thus, the unique objective is to minimize the number of \acp{pmu}, that results to be $d=11$. With this solution, we have only 3 \acp{ufc2}, as shown in Figure~\ref{fig:opa}(left). Therefore, the resulting fault localization resolution is not satisfactory. In particular, the entire Feeder~2 is included in the same \ac{ufc2} of a large portion of Feeder~1 (red lines cluster). In Case~B, optimization cost function weights \eqref{eq:cost_function_option2} are adopted to pursue the dual objective of minimizing the number $d$ of \acp{pmu} and maximizing the number $r$ of \acp{ufc2}. In this case, we obtain $d=12$, just one more with respect to Case~A, whereas $r=7$. Therefore, localization resolution, that we can measure by the number of \acp{ufc2}, is strongly augmented. In the simulation tests we adopt the results of Case~B. %In Figure~\ref{fig:Grid}, we can observe that two of the three nodes with Diesel generators result to be non monitored. Indeed, there is no requirement in the \ac{opa} about nodes with \ac{der}.

Simulations are realized on the Matlab platform. Measurements are simulated by adding noise to magnitude and phase of voltages and currents. Noises are assumed to be white and Gaussian, with the following standard deviations: $1.6\cdot10^{-3}\%$ and $4\cdot10^{-1}\%$ for voltage and current magnitudes, respectively; $5.1 \cdot 10^{-5}\si{\rad}$ and $5.8 \cdot 10^{-3}\si{\rad}$ for voltage and current phases, respectively. According to the capability of real \acp{pmu}, measures are sampled with a time step of \SI{20}{\milli \second}. Since measurements are in polar coordinates, a transformation to rectangular coordinates is applied \cite{milano:2016}. All these assumptions and noise standard deviations are the same of \cite{pignati2016fault}.

%\begin{table}
%		\centering
%		\caption{Measurements standard deviations.}
%		\label{tab:StandardDeviations}
%		\renewcommand{\arraystretch}{1.1}
%		\renewcommand{\tabcolsep}{0.25 cm}
%		\begin{tabular}{c c c c}
%		\hline \hline 
%		  \textbf{Voltage mag.}      & \textbf{Voltage angle}  & \textbf{Current mag.}  & \textbf{Current angle} \\  \hline 
%		  $1.6 \cdot 10^{-3} \% $ & $5.1 \cdot 10^{-5} ~\si{\rad}$ & $4 \cdot 10^{-1} \%$ &$5.8 \cdot 10^{-3} ~\si{\rad}$ \\
%        \hline \hline 
%\end{tabular}
%\end{table}

%%%%%%%%%%%%%%%%%%%%%%%%%%%%%%%%%%%%%%%%%%%%%%%%%
\section{Results} \label{sec:Results} 
%%%%%%%%%%%%%%%%%%%%%%%%%%%%%%%%%%%%%%%%%%%%%%%%
The \ac{fdla} has been tested on a set of 8 different short-circuit events, with fault resistances set to {\SI{100}{\ohm}}. For single-phase faults, both the cases with neutral connected to the ground directly and by Petersen coil. The latter will be indicated as 1-phase-p, the former with 1-phase-g. In the \ac{fdla}, detection threshold $th_w$ is equal to the 120\% of the maximum variation of $w^0$ registered in pre-fault conditions.

The results of a Monte Carlo analysis, with $100$ noise realizations for each fault, are reported in Table~\ref{tab:MCResults}. The table shows the number of occurrences of the following cases: (D-L) fault correctly detected and located; (D-nL) fault detected, but not correctly located; (n-DL) fault not detected, but correctly located; (nD-nL) fault not detected and not correctly located. 

It results that the performances of the proposed method are more than satisfactory. Indeed, all detection and localization percentages are close to 100\%. Worst results are registered for single-phase faults, especially in the case of neutral connection by Petersen coil. This is not surprising since, in these scenarios, fault currents are low and measurement noise may cause difficulties both in detection and in localization. We remark that, in all the cases, characterization has resulted to be correct. Moreover, perfect performances has been verified with lower levels of the measurement noises.

\begin{table}[t]
		\centering
		\caption{Results out of 100 simulations per scenario.}
		\label{tab:MCResults}
		\renewcommand{\arraystretch}{1}
		\begin{tabular}{l c c c c}
		\hline \hline	\vspace{-8pt} \\
		  \textbf{Fault} & \textbf{D-L} &\textbf{D-nL} &\textbf{nD-L} &\textbf{nD-nL} \\ \hline
		  {3-phase at $50\%$ of Line 4-5} &$100$ &$0$ &$0$ &$0$ \\ 
		  {1-phase-g at $50\%$ of Line 9-10} &$96$ &$4$ &$0$ &$0$ \\ 
		  {1-phase-p at $50\%$ of Line 9-10} &$90$ &$8$ &$2$ &$0$ \\
		  {2-phase at $50\%$ of Line 1-17} &$100$ &$0$ &$0$ &$0$ \\ 
		  {1-phase-g at $25\%$ of Line 7-8} &$99$ &$1$ &$0$ &$0$ \\ 
		  {1-phase-p at $25\%$ of Line 7-8} &$93$ &$5$ &$2$ &$0$ \\ 
		  {2-phase at $25\%$ of Line 13-14} &$100$ &$0$ &$0$ &$0$ \\ 
		  {3-phase at $75\%$ of Line 7-8} &$100$ &$0$ &$0$ &$0$ \\ 
		  {2-phase at $75\%$ of Line 9-10} &$100$ &$0$ &$0$ &$0$ \\ \hline\hline
\end{tabular}
\vspace{-10pt}
\end{table}

\begin{figure}[ht]
	\centering
	\includegraphics[width=1\columnwidth]{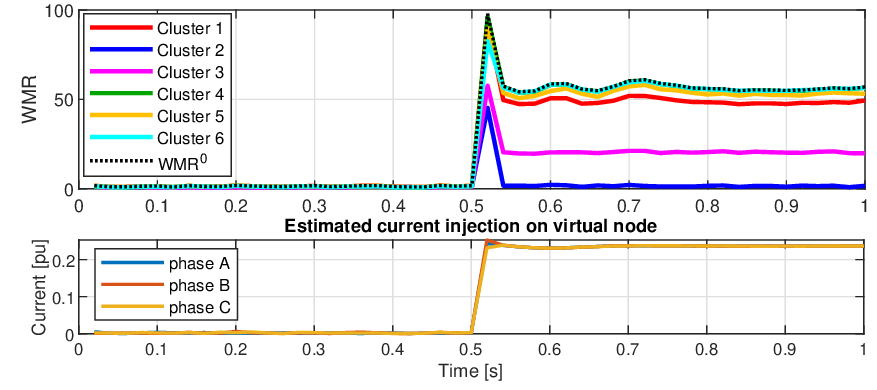}
    	\caption{\acp{wmr} (top) and estimated current injections at the virtual node (bottom) for a 3-phase fault at $50\%$ of Line 4-5 (Cluster 2).}
	\label{fig:Three-phaseFaultWMR}
\end{figure}

\begin{figure}[ht]
	\centering
	\includegraphics[width=1\columnwidth]{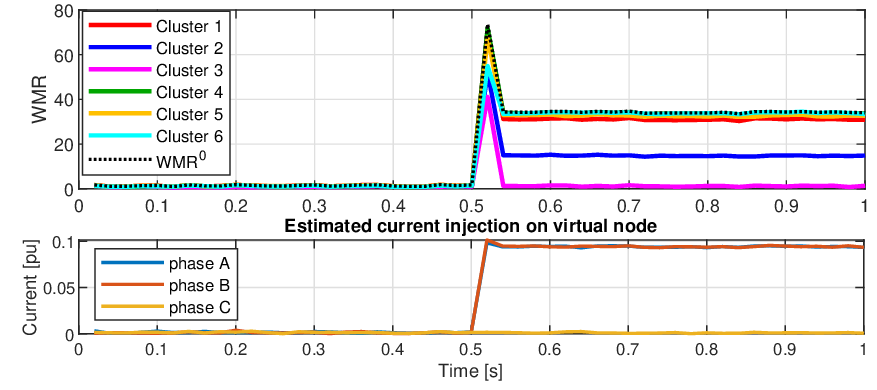}
    	\caption{\acp{wmr} (top) and estimated current injections at the virtual node (bottom) for a 2-phase fault at $75\%$ of Line 9-10 (Cluster 3).}
	\label{fig:Two-phaseFaultWMR}
	\vspace{-10pt}
\end{figure}

\begin{figure}[ht]
	\centering
	\includegraphics[width=1\columnwidth]{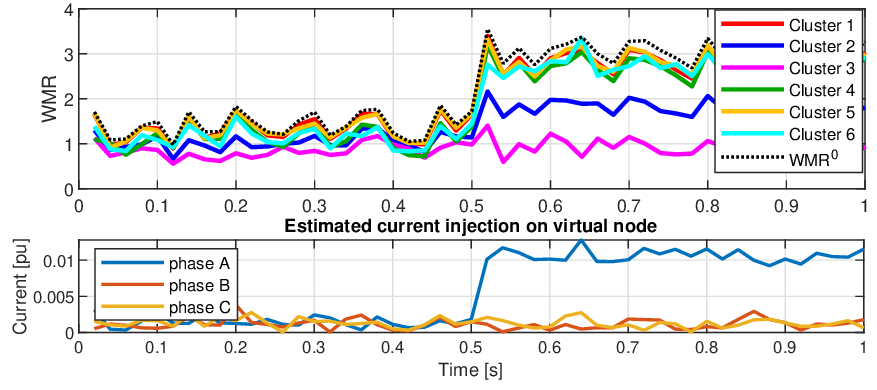}
    	\caption{\acp{wmr} (top) and estimated current injections at the virtual node (bottom) for a 1-phase-g fault at $25\%$ of Line 7-8 (Cluster 3).}
	\label{fig:Single-phaseFaultWMR}
	\vspace{-10pt}
\end{figure}

Figures~\ref{fig:Three-phaseFaultWMR}--\ref{fig:Single-phaseFaultWMR} show the detailed results of three specific fault occurrences, one 3-phase, one 2-phase and one 1-phase. In particular, the figures report the \acp{wmr} of seven \ac{se}, one (the $0$-th) associated to the grid without virtual nodes, used for detection, and six to the \acp{ufc}. The same colors reported in Fig.~\ref{fig:opa} are adopted to identify the different \acp{ufc}. Notice that the gray \ac{ufc} is constituted by the two primary transformers, an thus it is not considered.  
Moreover, the figures report the estimated current injections on the virtual node used to characterize the fault.

In all the cases, we can observe that at $\SI{0.5}{\second}$, when the faults occurs, the $0$-th \ac{wmr} suddenly increases allowing the detection. In the same time, the estimated current injections on the virtual node, corresponding to the faulted phases (phase A and B in the 2-phase case, and phase A in the 1-phase case), becomes different from zero, allowing the fault characterization. Finally, we can observe that the \ac{wmr} associated to the right cluster result to be the minimum immediately after the fault starting time, allowing a correct localization.

%%%%%%%%%%%%%%%%%%%%%%%%%%%%%%%%%%%%%%%%%%%%%%%%%%%%%%%%%%%%%%%%%%%
\section{Conclusions} \label{sec:Conclusions}
%%%%%%%%%%%%%%%%%%%%%%%%%%%%%%%%%%%%%%%%%%%%%%%%%%%%%%%%%%%%%%%%%%%
The objective of this paper was to determine the minimum number of \acp{pmu} required to realize a state estimation-based algorithm able to detect and localize faults in radial \acfp{adn}.
We discovered that, based on the positions of \acp{pmu} along the grid, faults are localizable within portions of the grid defined as clusters of lines. As such clusters are smaller, as higher is the fault localization resolution that we can obtain. As a consequence, an \acf{opa} has been developed, by which it is possible to simultaneously minimize the number of required \acp{pmu} and maximize the number of lines clusters. The approach has been successfully applied to a benchmark MV distribution network.

%------------------------------------------------------------------------------------
\vspace{.2cm}
\bibliographystyle{IEEEtran}

\end{document}